\documentclass[prd,showpacs,preprintnumbers]{revtex4}

\usepackage{graphicx}
\usepackage{dcolumn}
\usepackage{bm}
\usepackage{latexsym}
\usepackage{amsfonts}
\usepackage{amssymb}
\newcommand{\R }{ \mathop{R^{\mu}_{\ \nu}}^{(4)}  }
\newcommand{\Rs }{ \mathop{R}^{(4)}  }

\begin{document}

\preprint{KUCP0209 }

\title{Brane World Effective Action at Low Energies and AdS/CFT Correspondence}

\author{Sugumi Kanno}
\email{kanno@phys.h.kyoto-u.ac.jp}
 \affiliation{%
 Graduate School of Human and Environmental Studies, Kyoto University, Kyoto
 606-8501, Japan 
}%

\author{Jiro Soda}
\email{jiro@phys.h.kyoto-u.ac.jp}
\affiliation{
 Department of Fundamental Sciences, FIHS, 
Kyoto University, Kyoto 606-8501, Japan
}%

\date{\today}

\begin{abstract}
   A low energy iteration scheme to study  nonlinear gravity 
  in the brane world is developed.  As a result, we obtain 
  the brane world effective action at low energies.  
  The relation between the geometrical approach 
  and the approach using the AdS/CFT correspondence is also clarified.
   In particular, we  find {\sl   generalized dark radiation} as
    homogeneous solutions in our iteration scheme.  
    Moreover, the precise correspondence between the bulk geometry and 
    the brane effective action is established, which gives a holographic
    view of the brane world.
\end{abstract}

\pacs{98.80.Cq, 98.80.Hw, 04.50.+h}
\maketitle

\section{Introduction}

 The brane world scenario has been the subject of intensive investigation
  for the past few years. In particular, Randall and Sundrum   
  proposed a simple model
   where the four-dimensional brane with tension $\sigma$
   is embedded in the five-dimensional asymptotically anti-de Sitter (AdS) bulk
    with a curvature scale $l$. 
  The  model is described by the action
\begin{equation}
S = {1\over 2\kappa^2}\int d^5 x \sqrt{-g}\left( 
       {\cal R} + {12\over l^2} \right)  
       -\sigma \int d^4 x \sqrt{-h}
  + \int d^4 x \sqrt{-h} {\cal L}_{\rm matter} \ ,
\end{equation}
where ${\cal R}$ and $\kappa^2  $  are the scalar curvature and 
gravitational constant in five-dimensions, respectively. We  impose 
 $Z_2$ symmetry on this spacetime, with the brane at
 the fixed point ($y=0$ in the coordinate system used later). 
 Throughout this paper,
 $h_{\mu\nu}$ represents the induced metric on the brane.  
  They  showed that, in spite of the noncompact extra dimension,
   the gravity is localized on the brane at the  linearized level~\cite{RS,
   GT,GKR}. 
   Consequently, the conventional linearized Einstein equation 
   approximately holds at  scales large compared with
   the curvature scale $l$. 
  The cosmology in the context of this model  has also been investigated 
  enthusiastically~\cite{BWC0,BWC1,KJ0,BWC4,BWC5,
  BWC6,BWC8,BWC9,KJ,Muko,kodama,Bruck,Lan,Mar,Der,us}. 
   It turns out that there is a nonconventional
   quadratic term of the energy density at high energies 
  and, even in the low energy regime, 
  there exists  dark radiation  caused by the black hole
   in the bulk. This dark radiation component is also found in the
   cosmological perturbations~\cite{KJ2,Large}. 
   
   It is desirable to extend this understanding of gravity 
   in the brane world to  general nonlinear cases.  
   In order to understand  nonlinear gravity in the brane world scenario, 
   Shiromizu et al.  proposed an elegant geometrical  
   approach~\cite{SMS}.   
   They obtained  the  ``effective" four-dimensional equations 
\begin{equation}
 G^{(4)}_{\mu\nu} = 8\pi G_N T_{\mu\nu} + \kappa^4 \pi_{\mu\nu} 
                   - E_{\mu\nu}  \ , 
\end{equation}
where $T_{\mu\nu}$ is the energy-momentum tensor of matter,
$E_{\mu\nu}$ is the projection of the Weyl tensor  
$C_{y\mu y \nu} |_{y=0}$, $G_N$ denotes Newton's constant, and  
\begin{equation}
  \pi_{\mu\nu} = -{1\over 4}T_\mu^\lambda T_{\lambda\nu} 
  +{1\over 12}TT_{\mu\nu} + {1\over 8} g_{\mu\nu} \left( 
   T^{\alpha\beta} T_{\alpha\beta} -{1\over 3} T^2 \right) \ .
\end{equation}
Here the  relations
\begin{equation}
  \kappa^2 \sigma = {6\over l} \ ,\qquad   {\kappa^2 \over l} = 8\pi G_N
\end{equation}
have been assumed.
  From their effective equation of motion, one can see the ``electric" part
  of the Weyl tensor $E^\mu_{\ \nu}$ which characterizes the bulk geometry
   effects on the brane dynamics. Conversely, the matter on 
   the brane changes the bulk geometry, as can be seen from the equation 
   $\nabla_\mu E^\mu_{\ \nu} = \kappa^4 \nabla_\mu \pi^\mu_{\ \nu}$ 
   derived  through the Bianchi identity.
  However, it should be noted that it is by no means a closed system
   of equations. 
  We need to solve the bulk geometry  to determine  $E_{\mu\nu}$
   completely. 
  
  Unfortunately,  it is a formidable task to solve the five-dimensional 
  Einstein equation exactly. However, notice that typically the length
  scale of the internal space is $l\ll 0.1 $ mm. On the other hand, the
   usual astrophysical and cosmological phenomena take place at a scale 
   larger than this scale.
  Then we need only a low energy effective theory to analyze a variety
   of problems, for example,  the formation of the black hole, the propagation 
   of  the gravitational wave, the evolution of the cosmological perturbation,
   and so on. It should be stressed that  low energy does not necessarily
   imply  weak gravity.
   
  It has been suggested  that   gravity  on the brane at low energies can
   be understood through the  AdS/CFT correspondence~\cite{gubser}.
 From the correspondence, one can guess the effective equations of motion
  on the brane as
\begin{equation}
   G^{(4)}_{\mu\nu} = 8\pi G_N \left( T_{\mu\nu} + T^{CFT}_{\mu\nu} \right)
                   + \{ R^2 {\rm terms} \}   \ ,
\end{equation}
where the $R^2$ terms represent the higher order curvature terms and 
 $T^{CFT}$ denotes the energy-momentum tensor of the cutoff version 
 of  conformal field theory. However, no one knows what is the cutoff CFT.
  Moreover, it should be noted that the AdS/CFT  correspondence is a specific 
  conjecture. Indeed, originally, Maldacena conjectured that 
  the supergravity on $AdS_5 \times S^5$ is dual to the four-dimensional 
 ${\cal N}=4$ super Yang-Mills theory~\cite{malda}. 
 We should be careful not to use such a conjecture thoughtlessly.  
 Nevertheless,  the AdS/CFT correspondence seems to
 be related to the brane world model as has been  demonstrated 
 by several people~\cite{Haw,HS,haro1,haro2,KS2}.
 
  Since both the geometrical and AdS/CFT approaches seem 
  to have their own merit, 
  it would be beneficial to understand the mutual relationship.  
  Recently, Shiromizu and Ida tried to understand the AdS/CFT correspondence 
  from the  geometrical point of view~\cite{SI}.
   They argued that $\pi^\mu_{\ \mu}$ 
  corresponds to the trace anomaly of the cutoff CFT on the brane. 
  However, this  result is rather  paradoxical because there exists 
  no trace anomaly 
  in an odd dimensional  brane although  $\pi^\mu_{\ \mu}$ exists even in that 
  case.  Thus,  the more precise relation between the 
   geometrical and the AdS/CFT approaches remains to be understood. 

  In this paper, we  derive the effective four-dimensional theory 
 without using any concept of the AdS/CFT correspondence {\sl a priori}.  
 Thus, we  avoid using the vague concept of  cutoff CFT.  
 To solve the  five-dimensional equations of motion,  we  use a low energy 
 iteration scheme.  In particular, we  impose  the Dirichlet boundary 
 condition at the brane position, in contrast to the AdS/CFT approach where
 the Dirichlet boundary condition is imposed at infinity. We  also consider
 the ``constant" of the integration, i.e., homogeneous solutions, 
  carefully. As a consequence, we  show that the dark radiation can 
  be understood from the holographic point of view. 
  Moreover, the relation between the geometrical and AdS/CFT approaches
  is uncovered.
  The correspondence between the bulk geometry and the brane effective action
 is also explicitly found.

This paper is organized as follows. In Sec.II, we develop the iteration
 scheme to solve the Einstein equations at low energies.
In Sec.III,  we derive the brane effective action from the junction 
condition. We see the effective equation does not reduce to the conventional
 Einstein equation even in the low energy regime. This is due to
 the generalized  dark radiation found in this paper. 
 It is also found that the nonlocal part of the effective equation 
 is represented by the  energy momentum tensor
 with a trace part coinciding with the trace anomaly of CFT. 
 Sec.IV is devoted to the conclusion. 
 In the Appendix A, we analyze the $(d+1)$-dimensional case because the
 result is  qualitatively different from the five-dimensional case.

\section{Low Energy Iteration Scheme}

In the Gaussian normal coordinate system, the geometry of the brane world
is described by
\begin{equation}
ds^2 = dy^2 + g_{\mu\nu} (y,x^{\mu} ) dx^{\mu} dx^{\nu}  \ .
\end{equation}
Note that the brane is located at $y=0$ in this coordinate system. 
Then, the five-dimensional Einstein equation  becomes
\begin{eqnarray}
 & &K^\mu_{\ \nu ,y} - K K^\mu_{\ \nu} + {\R}   
                        = - {4\over l^2} \delta^\mu_\nu  
          + \kappa^2 \left({1\over 3}\sigma \delta^\mu_\nu
          +T^\mu_{\ \nu} -{1\over 3} \delta^\mu_\nu T \right)
                        \delta (y) \ ,\\
 & &K_{, y} - K^{\alpha  \beta} K_{\alpha  \beta} 
                           = - {4\over l^2}  
                           -{\kappa^2 \over 3}
                           \left( -4\sigma +T \right) \delta (y)         \ , \\
 & & \nabla_\nu K_{\mu }^{\  \nu}   - \nabla_\mu K    =  0  \ ,
\end{eqnarray}
where $\R$ is the curvature on the brane and $\nabla_\mu $ denotes the
 covariant derivative with respect to the metric $g_{\mu\nu}$.
One can read off the junction condition from the above equations as
\begin{equation}
\left[ K^\mu_\nu - \delta^\mu_\nu K \right] |_{y=0} 
    = {\kappa^2 \over 2}  \left( -\sigma \delta^\mu_\nu + T^\mu_\nu \right) \ .
\end{equation}
Recall that we are considering the $Z_2$ symmetric spacetime. 
Decomposing the extrinsic curvature into the traceless part and the trace part
\begin{equation}
  K_{\mu\nu} 
    = \Sigma_{\mu\nu }+ {1\over 4} h_{\mu\nu} K  \ , \quad
    K = - {\partial \over \partial y}\log \sqrt{-g}    \  , 
\end{equation}
we obtain the basic equations which hold in the bulk;
\begin{eqnarray}
 & & \Sigma^\mu_{\ \nu , y} - K \Sigma^\mu_{\ \nu} 
    = -\left[ \R - {1\over 4} \delta^\mu_\nu \Rs 
            \right]      \ ,     \\
 & &   {3\over 4} K^2 - \Sigma^\alpha_{\ \beta} \Sigma^\beta_{\ \alpha} 
    = \left[~ \Rs ~\right] + {12\over l^2}   \ ,  \\
 & &   K_{, y} -{1\over 4}K^2 - \Sigma^{\alpha \beta} \Sigma_{\alpha \beta} 
    = - {4\over l^2}     \ ,  \\
 & &   \nabla_\lambda \Sigma_{\ \mu }^{ \lambda}  
           -  {3\over 4} \nabla_\mu K = 0   \ .
\end{eqnarray}

The problem now is separated into two parts. First, we will solve
 the bulk equations of motion with the Dirichlet boundary condition
 at the brane, $g_{\mu\nu} (y=0 ,x^\mu ) = h_{\mu\nu} (x^\mu ) $.
 After that, the junction condition will be imposed at the brane.
 As it is the condition for the induced metric $h_{\mu\nu}$, it
 is naturally interpreted as the effective equations of motion
 for  gravity on the brane.
 
Along the normal coordinate $y$, the metric varies with a characteristic length
 scale $l$; $ g_{\mu\nu ,y} \sim g_{\mu\nu} /l$.  Denote the characteristic 
 length scale of the curvature fluctuation on the brane as $L$; then we have
  $ R \sim g_{\mu\nu} / L^2 $. For the reader's reference, let us take 
  $l=1$ mm, for example. Then, the relations (4) give the scale, 
  $\kappa^2 = (10^8 \ {\rm GeV})^{-3}$ and $\sigma = 1 \ {\rm TeV}^4$.  
  In this paper, we will consider the low energy
  regime in the sense that the energy density of matter, $\rho$, 
  on the brane is smaller than the brane tension, \i.e., $\rho /\sigma \ll 1$. 
  In this regime, a simple dimensional analysis
\begin{equation}
  {\rho \over \sigma} \sim {l {\kappa^2 \over l} \rho \over \kappa^2 \sigma}
    \sim \left({l\over L}\right)^2 \ll 1 
\end{equation}
implies that the curvature on the brane can be neglected compared with the
 extrinsic curvature at low energies. Here, we have used the relations (4)
 and Einstein's equation on the brane, $R\sim g_{\mu\nu}/L^2 \sim G_N \rho$.
 Thus, the anti-Newtonian or gradient expansion method used in the cosmological
 context is applicable to our problem~\cite{tomita,comer,salopek,soda}. 
The iteration scheme consists in writing the metric $g_{\mu\nu}$
 as a sum of local tensors built out of the induced metric on the
 brane, the number of gradients increasing with the order. 
Hence, we will seek the metric as a perturbative series
\begin{eqnarray}
    &&  g_{\mu\nu} (y,x^\mu ) =
  a^2 (y) \left[ h_{\mu\nu} (x^\mu) + g^{(1)}_{\mu\nu} (y,x^\mu)
      + g^{(2)}_{\mu\nu} (y, x^\mu ) + \cdots  \right]  \ , \\
   &&  g^{(i)}_{\mu\nu} (y=0 ,x^\mu ) =  0    \ ,
\end{eqnarray}
where $a^2 (y)= \exp (-2y/l) $ is extracted for reasons 
explained later and we put the Dirichlet boundary condition 
$g_{\mu\nu} (y=0, x) =  h_{\mu\nu} (x)$ at the brane. 
Other quantities are also expanded as
\begin{equation}
   K^\mu_{\ \nu} = K^{(0)\mu}_{\quad \ \nu}
            +K^{(1)\mu}_{\quad \ \nu}+K^{(2)\mu}_{\quad \ \nu}+\cdots  \ .
\end{equation}
Our scheme is different from the calculation usually performed
 in the AdS/CFT correspondence in that the Dirichlet boundary 
 condition is imposed not at infinity but at the finite
 point $y=0$, the location of the brane. Furthermore, we carefully keep the
  constants of integration, i.e., homogeneous solutions. These 
 homogeneous solutions are  ignored in the calculation of AdS/CFT 
 correspondence.  However, they play an important role in the
 brane world.

\subsection{Zeroth order}

At zeroth order,  we can neglect the curvature term. Then we have
\begin{eqnarray}
 & & \Sigma^{(0)\mu}_{\quad \  \nu , y} 
      - K^{(0)} \Sigma^{(0)\mu}_{\quad \  \nu} 
    = 0     \ ,  \\
 & &   {3\over 4} K^{(0)2} - \Sigma^{(0)\alpha}_{\quad \  \beta} 
    \Sigma^{(0)\beta}_{\quad \  \alpha} 
    =  {12\over l^2}   \ ,  \\
 & &   K^{(0)}_{, y} -{1\over 4}K^{(0)2} 
    - \Sigma^{(0)\alpha \beta} \Sigma^{(0)}_{\alpha \beta} 
    = - {4\over l^2}     \ ,   \\
 & &   \nabla_\lambda \Sigma^{(0)\lambda}_{\quad \ \mu }   
    -  {3\over 4} \nabla_\mu K^{(0)} = 0  \ .
\end{eqnarray}

 Equation (20) can be readily integrated into
\begin{equation}
    \Sigma^{(0)\mu}_{\quad \ \nu} = {C^\mu_{\ \nu} (x^\mu) \over \sqrt{-g} } 
                                 \ , \quad C^\mu_{\ \mu} =0  \ ,
\end{equation}
where $C^\mu_{\ \nu}$ is the integration ``constant." 
 Equation (23) also requires $C^\mu_{\ \nu |\mu} =0$.
It represents a radiationlike fluid on the brane. 
Although this  deserves further investigation, the calculation is complicated.
 Furthermore, this term is not relevant to our realistic universe because it 
 represents a strongly anisotropic universe. Indeed, as we see later, this
 term must vanish in order to satisfy the junction condition. 
 Therefore, we simply put $C^\mu_\nu =0$, hereafter. 
 Now, it is easy to solve the remaining equations. The result is
\begin{equation}
    K^{(0)} = {4\over l}    \ .
\end{equation}
Using the definition of the extrinsic curvature
\begin{equation}
     K^{(0)}_{\mu\nu} = - {1\over 2} {\partial \over \partial y} 
                               g^{(0)}_{\mu\nu}     \   ,
\end{equation}
we get the zeroth order metric  as
\begin{equation}
 ds^2 = dy^2 +  a^2 (y) h_{\mu\nu}(x^\mu ) dx^\mu dx^\nu  \ , \quad
    a(y)  = e^{-2{y\over l}}    \ ,
\end{equation}
where  the tensor $h_{\mu\nu}$ is  the induced metric on the brane.

\subsection{First order}

The next order solutions are obtained by taking into account the 
terms neglected at zeroth order. 
At  first order,  Eqs. (12) - (15) become
\begin{eqnarray}
 & & \Sigma^{(1) \mu}_{\quad \  \nu , y} 
      - {4\over l} \Sigma^{(1) \mu}_{\quad \  \nu} 
    = -\left[ \R - {1\over 4} \delta^\mu_\nu \Rs \right]^{(1)} \ , \\
 & &   {6 \over l} K^{(1)}  = \left[~ \Rs ~\right]^{(1)}   \  ,\\
 & &   K^{(1)}_{, y} -{2\over l}K^{(1)} = 0   \ ,\\
 & &   \Sigma_{\mu \ \ |\lambda}^{(1)\ \lambda}   
              -  {3\over 4} K^{(1)}_{|\mu} = 0 \ .
\end{eqnarray}
where the superscript $(1)$ represents the order of the derivative expansion
 and $|$ denotes the covariant derivative with respect to
 the metric $h_{\mu\nu}$.
Here, $[R^\mu_\nu ]^{(1)} $ means that the curvature is approximated by
 taking the Ricci tensor of $a^2 h_{\mu\nu} $ in place of 
 $R^{(4)\mu}_{\quad \nu}$. 
 It is also convenient to write it in terms of the Ricci
  tensor of $h_{\mu\nu}$, denoted $R^\mu_\nu (h)$.
 
Substituting the zeroth order metric into $R^{(4)}$, we obtain
\begin{equation}
   K^{(1)} = {l \over 6a^2} R(h)    \ .
\end{equation}
Hereafter, we omit the argument of the curvature for simplicity. 
Simple integration of Eq.(28) also gives the traceless part
 of the extrinsic curvature as
\begin{equation}
  \Sigma^{(1)\mu}_{\ \nu} =  {l\over 2 a^2 }
      ( R^\mu_{\ \nu}  - {1\over 4} \delta^\mu_\nu R )  
        + {\chi^{\mu}_{\ \nu} (x) \over a^4}  \ ,
\end{equation}
where the homogeneous solution satisfies the constraints 
$\chi^{\mu}_{\ \mu} =0 $ and $\chi^{\mu}_{\ \nu|\mu}=0 $.
 As we see later, this term  corresponds to  dark 
 radiation at this order.  
The metric can be obtained as
\begin{equation}
  g^{(1)}_{\mu\nu} = -{l^2 \over 2} \left( {1\over a^2}-1 \right) 
    \left( R_{\mu\nu}  - {1\over 6} h_{\mu\nu} R \right) 
    -{l \over 2}\left( {1\over a^4} -1 \right) \chi_{\mu \nu} \ ,
\end{equation}
where we have imposed the boundary condition, 
$g^{(1)}_{\mu\nu} (y=0, x^\mu ) =0 $. 
 This $\chi$ field is essential to understanding the origin of the 
 dark radiation from the holographic point of view.

\subsection{Second order}

In this subsection, we do not include the $\chi$ field 
because it complicates the  equations. This is a consistent truncation 
procedure. Of course, we have  calculated the second order solutions 
with the contribution of the $\chi$ field.
 They include  terms such as $\chi^\mu_{\ \nu}\chi^\nu_{\ \mu}$, etc. 
 
At  second order, the basic equations become 

\begin{eqnarray}
   & & \Sigma^{(2) \mu}_{\quad \   \nu , y} 
        - {4\over l} \Sigma^{(2) \mu}_{\quad \  \nu} 
    = -\left[ \R - {1\over 4} \delta^\mu_\nu \Rs \right]^{(2)}
          + K^{(1)} \Sigma^{(1) \mu}_{\quad \  \nu}    \ ,\\
   &  & K^{(2)} = {l\over 6} \left[-{3\over 4} K^{(1) 2} 
      +  \Sigma^{(1)\alpha}_{\quad \  \beta} 
      \Sigma^{(1)\beta}_{\quad \  \alpha} 
      + \left[~ \Rs ~\right]^{(2)}  \right]
                                   \ , \\
   & & K^{(2)}_{, y} -{2\over l}K^{(2)} = {1\over 4} K^{(1)2}
     + \Sigma^{(1)\alpha \beta} \Sigma^{(1)}_{\alpha \beta} 
                                   \ , \\
   & & \Sigma_{\mu \ \ |\lambda}^{(2)\ \lambda}   -  {3\over 4} K^{(2)}_{|\mu} 
    + \Gamma^{(1)\alpha }_{\lambda \alpha} \Sigma^{(1)\lambda}_{\ \mu}
       - \Gamma^{(1)\lambda }_{\alpha\mu} \Sigma^\alpha_{\ \lambda} =0 
                                   \ .
\end{eqnarray}

Substituting the solution up to  first order into the Ricci tensor
 and picking up the second order quantities, we obtain
\begin{eqnarray}
   \left[\R -{1\over 4} \delta^\mu_\nu \Rs \right]^{(2)}  
       &=& {l^2\over 2}\left({1\over a^4}-{1\over a^2}\right) \left[ 
   R^\mu_{\ \alpha} R^\alpha_{\ \nu} -{1\over 6} R R^\mu_{\ \nu} 
   -{1\over 4} \delta^\mu_\nu (R^\alpha_{\ \beta} R^\beta_{\ \alpha}
         - {1\over 6} R^2)      \right.  \nonumber \\
     & & \left. \qquad  -{1\over 2} ( R^{\alpha\mu}_{\ \ \ |\nu\alpha}
                    + R^{\alpha}{}_{\nu}{}^{|\mu}{}_{|\alpha} )
 +{1 \over 3} R^{|\mu}_{\ |\nu}  +{1\over 2} \Box R^\mu_{\ \nu} 
              -{1\over 12} \delta^\mu_\nu \Box R   \right]   \ , \\
    \left[~ \Rs ~\right]^{(2)}  &=& {l^2\over 2}
    \left({1\over a^4}-{1\over a^2}\right)
    \left[   R^\alpha_{\ \beta} R^\beta_{\ \alpha}
         - {1\over 6} R^2   \right]  \ , 
\end{eqnarray}
where we have used the formula $\delta R_{\mu\nu} 
 = 1/2 [ \delta g^\alpha_{\ \mu |\nu\alpha} +\delta g^\alpha_{\ \mu |\nu\alpha}
 - \delta g_{|\mu\nu} -\delta g_{\mu\nu | \alpha }^{\qquad|\alpha} ]$.
Using the above formula, the trace part is deduced algebraically as
\begin{equation}
  K^{(2)}  = {l^3 \over 8 a^4} \left( R^\alpha_{\ \beta} R^\beta_{\ \alpha} 
            - {2\over 9}  R^2 \right) -{l^3 \over 12 a^2} 
            \left( R^\alpha_{\ \beta} R^\beta_{\ \alpha} 
            - {1\over 6}  R^2 \right)   \ .
\end{equation}
 By integrating the equation for the traceless part, we have
\begin{equation}
  \Sigma^{(2)\mu}_{\ \ \  \nu}  = -{l^2\over 2}\left( {y\over a^4} 
       + {l\over 2a^2} \right)
   {\cal S}^\mu_{\ \nu}  - {l^3 \over 24 a^2} \left( R R^\mu_{\ \nu} 
            - {1\over 4}  \delta^\mu_\nu R^2 \right) 
            + {l^3 \over a^4}t^\mu_{\ \nu}   \ ,
\end{equation}
where we have defined the quantity
\begin{eqnarray}
  {\cal S}^\mu_{\ \nu} &=& R^\mu_{\ \alpha} R^\alpha_{\ \nu}
             -{1\over 3} R R^\mu_{\ \nu} 
         -{1\over 4} \delta^\mu_\nu (R^\alpha_{\ \beta} R^\beta_{\ \alpha}
         - {1\over 3} R^2)  \nonumber \\ 
    & & \qquad     -{1\over 2} \left( R^{\alpha\mu}{}_{|\nu\alpha}
                    + R^{\alpha}{}_{\nu}{}^{|\mu}{}_{|\alpha}  
              -{2\over 3} R^{|\mu}{}_{|\nu}  - \Box R^\mu_{\ \nu} 
              +{1\over 6} \delta^\mu_\nu \Box R \right)   \ ,
\end{eqnarray}
which is transverse and traceless,
\begin{equation}
  {\cal S}^\mu_{\ \nu | \mu} =0  \ ,  \quad   {\cal S}^\mu_{\ \mu} = 0   \ .
\end{equation}
It is also useful to notice that this tensor can be derived from
\begin{equation}
   \delta \int d^4 x \sqrt{-h} {1\over 2} \left[ 
       R^{\alpha\beta} R_{\alpha\beta} -{1\over 3}R^2 \right] 
       = \int d^4 x \sqrt{-h} {\cal S}_{\mu\nu}
                         \delta g^{\mu\nu}     \ .
\end{equation}
The homogeneous solution $t^\mu_{\ \nu} $
 must be determined so that 
 Eq.(38) holds. To be more precise, we must solve the constraint equation
\begin{equation}
    t^\mu_{\ \nu |\mu} - {1\over 16} R^\alpha_{\ \beta} R^\beta_{\ \alpha|\nu}
     + {1\over 48} R R_{|\nu} - {1\over 24} R_{|\lambda} R^\lambda_{\ \nu}  
                                             =0  \ .
\end{equation}
As one can see immediately,
 there are ambiguities in integrating this equation. 
Indeed, there are two types of covariant local tensor whose
 divergences vanish:
 \begin{eqnarray}
     {\cal H}^\mu_{\ \nu} &=& R^\mu_{\ \alpha} R^\alpha_{\ \nu} 
         -{1\over 4} \delta^\mu_\nu R^\alpha_{\ \beta}R^\beta_{\ \alpha}
        -{1\over 2} \left( R^{\alpha\mu}{}_{|\nu\alpha}
                    + R^{\alpha}{}_{\nu}{}^{|\mu}{}_{|\alpha}  
                     - \Box R^\mu_{\ \nu} 
              - {1\over 2} \delta^\mu_\nu \Box R \right)  \ , \\
   {\cal K}^\mu_{\ \nu} &=&  R R^\mu_{\ \nu} 
         -{1\over 4} \delta^\mu_\nu R^2     
              - R^{|\mu}_{\ |\nu}  + \delta^\mu_\nu \Box R \ .
 \end{eqnarray}
 These terms come from the variation of the action
\begin{eqnarray}
  & & \delta \int d^4 x \sqrt{-h} {1\over 2}R^{\alpha\beta} R_{\alpha\beta} 
  = \int d^4 x \sqrt{-h} {\cal H}_{\mu \nu} \delta g^{\mu\nu}    \ , \\
  & & \delta \int d^4 x \sqrt{-h} {1\over 2} R^2 
   = \int d^4 x \sqrt{-h} {\cal K}_{\mu \nu} \delta g^{\mu\nu}  \ , 
\end{eqnarray}
respectively. Notice that ${\cal S}^\mu_{\ \nu} = {\cal H}^\mu_{\ \nu}
 - {\cal K}^\mu_{\ \nu} /3 $. Thanks to the Gauss-Bonnet topological 
 invariant, we do not need to consider the Riemann squared term.  
 In addition to these local tensors, we have to take into account
  the nonlocal tensor $\tau^\mu_{\ \nu}$ 
  with the property $\tau^\mu_{\ \nu|\mu}=0$. Thus, we get
\begin{equation}
 t^\mu_{\ \nu} 
       = {1\over 32}\delta^\mu_\nu 
          \left( R^\alpha_{\ \beta} R^\beta_{\ \alpha} 
            - {1\over 3}  R^2 \right) 
            +{1\over 24} \left( R R^\mu_{\ \nu} 
         -{1\over 4} \delta^\mu_\nu R^2    \right)
        + \tau^\mu_{\ \nu}  + \alpha {\cal S}^\mu_{\ \nu} 
        + {\beta \over 3} {\cal K}^\mu_{\ \nu}  \ , 
\end{equation}
where the constants $\alpha$ and $\beta$ represents the freedom of 
the gravitational wave in the bulk. 
 The condition  $t^\mu_{\ \mu} =0$ leads to
\begin{equation}
  \tau^\mu_{\ \mu}  
     =-{1\over 8} \left( R^\alpha_{\ \beta} R^\beta_{\ \alpha} 
            - {1\over 3}  R^2 \right) - \beta \Box R      \ .
\end{equation}
The quantity $\tau^\mu_{\ \nu}$ cannot 
be written in the local covariant form. 
Hence, this part is interpreted as the CFT in the context of the 
 AdS/CFT correspondence.

\subsection{$n$ th order}

In principle, we can continue our analysis up to a desired order
 using the following recursive formulas: 
\begin{eqnarray}
    & & \Sigma^{(n) \mu}_{\quad \   \nu }  
    = - {1\over a^4} \int dy a^4 \left\{
    \left[ \R - {1\over 4} \delta^\mu_\nu \Rs \right]^{(n)}
        -\sum_{p=1}^{n-1} K^{(p)} 
        \Sigma^{(n-p) \mu}_{\quad \quad  \nu} \right\} 
                                    \ , \\
   &  & K^{(n)} = {l\over 6} \sum_{p=1}^{n-1}
         \left[-{3\over 4} K^{(p)}  K^{(n-p)}
      +  \Sigma^{(p)\alpha}_{\quad \  \beta} 
      \Sigma^{(n-p)\beta}_{\quad \quad  \alpha} 
      + \left[~ \Rs ~\right]^{(n)}\right]
        \ , \\
   & & K^{(n)}_{, y} -{2\over l}K^{(n)} 
   =  \sum_{p=1}^{n-1} \left\{ {1\over 4} K^{(p)}K^{(n-p)}
     + \Sigma^{(p)\alpha \beta} \Sigma^{(n-p)}_{\alpha \beta} 
                      \right\} \ , \\
   & & \Sigma_{\mu \quad \ |\lambda}^{(n)\ \lambda}  
    -  {3\over 4} K^{(n)}_{|\mu} 
    + \sum_{p=1}^{n-1} \left\{ \Gamma^{(p)\alpha }_{\lambda \alpha} 
    \Sigma^{(n-p)\lambda}_{\quad\quad \mu}
       - \Gamma^{(p)\lambda }_{\alpha\mu} 
       \Sigma^{(n-p)\alpha}_{\quad\quad \lambda} \right\}=0
                                            \ . 
\end{eqnarray}
 As one can see from Eq.(53),  homogeneous solutions will appear
 at each order. However,  we note that the subtlety discussed 
 in the second order calculations
 never occurs in the higher order calculations.
 The existence of the infinite series is a manifestation of the nonlocality
 of the brane model. Therefore, we have two kinds of nonlocality on the 
 brane. One is the nonlocality associated with homogeneous solutions and
 the other is the infinite series which is the reflection of the extent 
 in the $y$ direction.

\section{Effective Equations and Effective Action}

 Now consider the consequences of the junction condition (10).
 The  findings in this section are the following. We find the generalized dark
 radiation $\chi^\mu_{\ \nu}$. The quadratic correction $\pi^\mu_{\ \nu}$ is
 identified with $P^\mu_{\ \nu}$ which is the local tensor defined later. 
 The relation between the geometrical approach and the AdS/CFT approach is 
 revealed. The brane effective action is obtained and the corresponding bulk 
 geometry is given explicitly.

\subsection{Zeroth order}

 From the zeroth order solution, we obtain
\begin{equation}
   \left[ K^{(0)\mu}_{\quad\quad\nu} 
   - \delta^\mu_\nu K^{(0)} \right] |_{y=0}
    = -{3 \over l} \delta^\mu_\nu
    = - {\kappa^2 \over 2} \sigma \delta^\mu_\nu  \ .
\end{equation}
Then we get the well known relation $\kappa^2 \sigma = 6/l$.
Here, we will assume that this relation holds exactly. 
 It is apparent that $C^\mu_{\ \nu}$ is not allowed to exist.

\subsection{First order}

Let us focus on the role of  $\chi^\mu_{\ \nu}$ in this part.
 At this order, the junction condition can be written as
\begin{equation}
  \left[ K^{(1)\mu}_{\quad\quad \nu} 
  - \delta^\mu_\nu K^{(1)} \right] |_{y=0}
  = {l\over 2} \left( R^\mu_{\ \nu} -{1\over 2} \delta^\mu_\nu R \right)
    + \chi^\mu_{\ \nu}
  = {\kappa^2 \over 2} T^\mu_{\ \nu}  \ .
\end{equation}
Using the solutions obtained in the previous section and the formula
\begin{equation}
   E^\mu_{\ \nu} = K^\mu_{\ \nu ,y} - \delta^\mu_\nu K_{,y}
    - K^\mu_{\ \lambda } K^\lambda_{\ \nu }  
    +\delta^\mu_\nu K^\alpha_{\ \beta } K^\beta_{\ \alpha }
      - {3\over l^2} \delta^\mu_\nu   \ ,
\end{equation}
we  calculate the projective Weyl tensor as 
$E^{(1)\mu}_{\quad\ \nu} =2/l \chi^\mu_{\ \nu} $.
Then we obtain the effective Einstein equation
\begin{equation}
 R^\mu_{\ \nu} -{1\over 2} \delta^\mu_\nu R  
 ={\kappa^2 \over l} T^\mu_{\ \nu} - E^{(1)\mu}_{\quad\ \nu} \ .
\end{equation}
At this order, we do not have the conventional Einstein equations. 
 Recall that  the dark radiation exists even 
 in the low energy regime. Indeed, the low energy effective Friedmann
 equation becomes
\begin{equation}
  H^2 = {8\pi G_N \over 3} \rho + {C \over a_0 (t)^4}  \ ,
\end{equation}
where $a_0 (t)$, $H$, and $C$ denote the scale factor on the brane, 
 the Hubble constant, and a constant, respectively.  
This equation can be obtained from Eq.(60) by
 imposing the maximal symmetry on the spatial part of the brane world.
Hence, we observe that $\chi^\mu_{\ \nu}$ is the generalization of 
the dark radiation found in the cosmological context. 

The conventional Einstein gravity can be recovered 
  on the brane at this order when we adopt the boundary condition for which
  $\chi^\mu_{\ \nu}$ vanish. 

\subsection{Second order}

In this subsection, we assume $\chi^\mu_{\ \nu}=0$. 
Up to the second order, the junction condition gives
\begin{equation}
 R^\mu_{\ \nu} -{1\over 2} \delta^\mu_\nu R 
             +2 l^2 \left[ \tau^\mu_{\ \nu} 
             +\left(\alpha-{1\over 4}\right)  {\cal S}^\mu_{\ \nu}
             +{\beta\over 3} {\cal K}^\mu_{\ \nu} \right] 
            = {\kappa^2 \over l} T^\mu_{\ \nu}    \ .
\end{equation}
Let us try to arrange the terms so as to reveal the geometrical
 meaning of the above equation. 
 We can calculate the projective Weyl  tensor as
\begin{equation}
  E^{(2)\mu}_{\quad \ \nu} 
  = l^2  \left[P^\mu_{\ \nu} +2\tau^\mu_{\ \nu} 
  +2 \left(\alpha -{1\over 4}\right) {\cal S}^\mu_{\ \nu} 
    + {2\over 3} \beta {\cal K}^\mu_{\ \nu} \right]     \ .
\end{equation}
Substituting this expression into Eq.(62) yields our main result
\begin{equation}
  G^{(4)}_{\mu\nu} = {\kappa^2 \over l} T_{\mu\nu}   
  +l^2 P_{\mu\nu}  - E^{(2)}_{\mu\nu} \ ,
\end{equation}
where
\begin{equation}
   P^\mu_{\ \nu } = -{1\over 4}R^\mu_{\ \lambda} R^\lambda_{\ \nu}  
              +{1\over 6}R R^\mu_{\ \nu} 
         +{1\over 8}\delta^\mu_\nu R^\alpha_{\ \beta} R^\beta_{\ \alpha}
            - {1\over 16} \delta^\mu_{\ \nu} R^2  \ .
\end{equation}
Notice that $E^\mu_{\ \nu}$ contains the nonlocal part and the free 
parameters $\alpha$ and $\beta$. 
 On the other hand, $P^\mu_{\ \nu}$ is determined locally. 
 If we define $T_{\mu\nu}^{\rm CFT}= -2l^3/\kappa^2 \tau_{\mu\nu} $, 
we can write
\begin{equation}
     G^{(4)}_{\mu\nu} = 8\pi G_N \left( 
     T_{\mu\nu} + T_{\mu\nu}^{\rm CFT} \right)
         -2 l^2 \left( \alpha -{1\over 4} \right) {\cal S}_{\mu\nu} 
         - {2 l^2 \over 3} \beta {\cal K}_{\mu\nu}  \ .
\end{equation}
 It is possible to use the result of CFT at this point.  
 For example, we can choose the ${\cal N}=4$ super Yang-Mills theory
 as the conformal matter. In that case, we  simply put $\beta =0$. 
 This is the
 way the AdS/CFT correspondence comes into the brane world scenario. 
Thus we get an explicit relation between the geometrical approach and the
 AdS/CFT approach.
 One can see the relationship in a different way.  
Within the accuracy we are considering,  
we can get $ P^\mu_{\ \nu } =  \pi^\mu_{\ \nu }$ using the lowest order 
equation 
$R^\mu_{\ \nu} = {\kappa^2 /l} (T^\mu_{\ \nu} - 1/2 \delta^\mu_\nu T) $.
 Hence,  we can rewrite Eq.(64) as 
\begin{equation}
  G^{(4)}_{\mu\nu} = 8\pi G_N T_{\mu\nu} + \kappa^4 \pi_{\mu\nu} 
       -  E^{(2)}_{\mu\nu}  \ .
\end{equation}
 Now, the similarity between Eq.(2) and Eq.(67) is apparent. 
However, we note that our Eq.(67) is a closed system
 of equations provided that the specific conformal field theory
 is chosen.

Now we can read off the effective action as
\begin{eqnarray}
   S_{\rm eff}  &=& {1\over 16\pi G_N } \int d^4 x \sqrt{-h} R
                    + S_{\rm matter} + S_{CFT}   \nonumber \\
    & &  \quad +{(\alpha -{1\over 4} )l^2 \over 16 \pi G_N}
     \int d^4 x \sqrt{-h}
           \left[R^{\mu\nu} R_{\mu\nu}
                   - {1\over 3}  R^2 \right] 
      +{\beta l^2 \over 48 \pi G_N} \int d^4 x \sqrt{-h} R^2  \ ,
\end{eqnarray}
where we have used the relations (45), (49) and (50) and we denoted the 
nonlocal effective action constructed from $\tau^\mu_{\ \nu}$ as $S_{\rm CFT}$. 
 This effective action corresponds to the bulk geometry given by the metric
\begin{eqnarray}
  g_{\mu\nu} (y,x^\mu) &=& a^2 \left[ 
          h_{\mu\nu} 
       - {l^2 \over 2 } \left( {1\over a^2}-1 \right)
       \left( R_{\mu\nu} -{1\over 6}h_{\mu\nu}R^2 \right)  
       +{l^3 \over 4}\left( {y\over a^4}-{l\over 4a^4}
           +{l\over a^2} -{3l\over 4} \right)  {\cal S}_{\mu\nu}  
                                             \right. \nonumber\\
    & &   \qquad    -{l^4 \over 2}\left( {1\over a^4}-1 \right) 
          \left( \tau_{\mu\nu} +\alpha {\cal S}_{\mu\nu}
                          +{\beta\over 3} {\cal K}_{\mu\nu} \right)
                                          \nonumber\\
    & &  \qquad + {l^4 \over 8} \left( {1\over a^4}-1 \right)
       \left( R_{\mu\lambda} R^\lambda_{\ \nu}  
              -{1\over 2}R R_{\mu\nu}  
              -{1\over 4} h_{\mu\nu} R^\alpha_{\ \beta} R^\beta_{\ \alpha}
            +{5\over 36} h_{\mu\nu} R^2 \right)   \nonumber\\
     & & \qquad \left.  -   {l^4 \over 4} \left( {1\over a^2}-1 \right)
       \left( R_{\mu\lambda} R^\lambda_{\ \nu}  
              -{1\over 2} R R_{\mu\nu}  
              -{1\over 12}  h_{\mu\nu} R^\alpha_{\ \beta} R^\beta_{\ \alpha}
            +{1\over 12} h_{\mu\nu} R^2 \right)   \right]  \ .
\end{eqnarray}
This gives the holographic view of the bulk geometry. The bulk geometry
 can be reconstructed provided the additional knowledge of the
 nonlocal component $\tau^\mu_{\ \nu}$ and the constants $\alpha$ and $\beta$
  is available.
 Both represent the effects of the bulk geometry, which is apparent
  because they appear in the projective Weyl tensor $E^\mu_{\ \nu}$.

\section{Conclusion}

We  developed a low energy iteration scheme for solving the equations
 of the brane world model. Using this formalism, 
 we  explicitly identified the low energy equations of motion describing
   gravity on the brane. The result should be useful in the investigation of
  various phenomena occurring in the brane world because it can treat  
  nonlinear gravity as far as  $l^2 R(h)\ll 1$. 
  
  Our work was motivated by two important approaches,
  the geometric approach and the AdS/CFT approach. 
  In fact, one of the purposes of
   this work was to clarify   the relation  between these approaches. 
  In  previous work, the trace anomaly of the CFT is identified with 
 $\pi^\mu_{\ \mu} $. 
 This interpretation is rather paradoxical because, in the odd 
 dimensional brane, no trace anomaly exists although $\pi^\mu_{\ \mu} $ exists 
 (see the Appendix A). 
 We  clarified this point by calculating the Weyl tensor. It turned out 
 that, irrespective of dimensions, 
 $\pi^\mu_{\ \nu}$ corresponds to $P^\mu_{\ \nu}$
  at low energies. In the case of the four-dimensional brane, the trace part
  of $P^\mu_{\ \nu}$ accidentally coincides with the trace anomaly of the
  CFT.

  We  found two kinds of  nonlocality observed on the brane. 
  One of them is encoded in the homogeneous solutions,
   and the other is found as an infinite series of the gradient expansion 
   of our scheme. Thus even  when we truncate the series at the second order,
    the knowledge of the  homogeneous solutions is  needed to solve 
    the problem. Indeed, there are two homogeneous solutions because 
    the system is described by a second order differential
  equation. One is used to satisfy the Dirichlet boundary condition at the 
  brane. The other appears as the ``dark" effects on the brane at each order
   of expansion in our scheme. At the zeroth order, $C^\mu_{\ \nu} $ appears.
   However, this term must vanish from  consistency.
    It is at the first order that the generalized dark radiation term 
   $\chi^\mu_{\ \nu} $  appeared. This term reduces to the
   dark radiation in the effective Friedmann equation under the assumption
   of  homogeneity. Note that
  it is possible to put $\chi^\mu_{\ \nu} =0$ if one prefers. 
  This can be achieved by putting the black hole mass to zero in
  the cosmological case. As for the general cases,  further consideration
  is needed. 
   At  second order, we  deduced the nonlocal component
   $\tau^\mu_{\ \nu} $ from the homogeneous solution. At this time, 
  it is far from possible to put  $\tau^\mu_{\ \nu} =0$ without losing 
   consistency. 
  We must treat it as an integro-differential equation~\cite{mukoh}
   or  coupled equations. 
 
   Needless to say, the  ambiguity of the effective action comes from 
    the variety of the bulk geometry. We have given the explicit
   correspondence between the effective action and the bulk metric,
   which could give a holographic view of the brane world. 
   Of course, this ambiguity should be fixed by  proper consideration of
   the boundary condition  in the bulk~\cite{wiseman}.
   Once the boundary condition is determined, we can attack various 
   astrophysical  and cosmological problems.
    
  As an application of our results, it would be interesting to consider the
  nature of the gravitational wave in the brane world. It is also
  important to investigate the quantum brane world from this point of view.
  In particular, we will apply our formalism to the inflation model driven 
  by a bulk scalar field~\cite{KKJ,SH}.  
  The analysis of the present paper can be 
  extended to the two brane system. In particular, the low energy dynamics
   of the radion could be treated by means of our method~\cite{kanno}.
  We will also study  more general models like the 
  Horava-Witten model in the future.

\begin{acknowledgements}
We would like to thank the participants of the YITP workshop
 YITP-W-01-15 on ``Braneworld - Dynamics of Spacetime Boundary"
  for fruitful discussions. This work was supported in part by
  Monbukagakusho Grant-in-Aid No.14540258.
\end{acknowledgements}

\appendix

\section{$(d+1)$-dimensional results  ($d\neq 4$ )} 

 The qualitative consequences of the low energy expansion scheme depend
 on the dimensions. Hence, for completeness, we  investigated 
 the ($d+1$)-dimensional problem. 
 
 We obtain basic equations which hold in the bulk as follows: 
\begin{eqnarray}
 & & \Sigma^\mu_{\ \nu , y} - K \Sigma^\mu_{\ \nu} 
        = -\left[ \mathop{R^{\mu}_{\ \nu}}^{(d)} - {1\over d} \delta^\mu_\nu 
        \mathop{R}^{(d)}             \right]      \ ,     \\
 & & {d-1\over d} K^2 - \Sigma^\alpha_{\ \beta} \Sigma^\beta_{\ \alpha} 
        = R^{(d)} + {d(d-1)\over l^2}   \ ,  \\
 & & K_{, y} -{1\over d}K^2 - \Sigma^{\alpha \beta} \Sigma_{\alpha \beta} 
        = - {d\over l^2}     \ ,  \\
 & & \nabla_\lambda \Sigma_{\mu }^{\ \lambda}  
           -  {d-1\over d} \nabla_\mu K = 0   \ .
\end{eqnarray}

 Since the calculations are similar to those in the $d=4$ case, 
 we simply write down  the results in the following subsections. 
\subsection{Zeroth order}

At zeroth order, we have
\begin{equation}
    K^{(0)} = {d\over l}    \ .
\end{equation}
  The zeroth order metric is given by
\begin{equation}
 ds^2 = dy^2 +  a^2 (y) h_{\mu\nu}(x^\mu ) dx^\mu dx^\nu  \ ,\ \ 
    a(y) = e^{-2{y\over l}}    \ .
\end{equation}

\subsection{First order}

 At  first order, the solutions are
\begin{eqnarray}
   K^{(1)} &=& {l \over 2(d-1)a^2} R(h)    \ ,  \\
  \Sigma^{(1)\mu}_{\quad \   \nu } &=&  {l\over (d-2) a^2 }
      \left( R^\mu_{\ \nu}  - {1\over d} \delta^\mu_\nu R \right) 
      + {\chi^\mu_\nu \over a^d}   \ .
\end{eqnarray}
where $\chi^\mu_\nu$ is a homogeneous solution which satisfies 
$\chi^{\mu}_{\ \mu} =0 $ and $\chi^{\mu}_{\ \nu|\mu}=0 $. This 
corresponds to the dark radiation at this order.
  The metric can be deduced as
\begin{equation}
  g^{(1)}_{\mu\nu} = -{l^2 \over d-2} \left( {1\over a^2}-1 \right) 
    \left( R_{\mu\nu}  - {1\over 2(d-1)} h_{\mu\nu} R \right) 
    -{2l \over d}\left( {1 \over a^d} - 1 \right) \chi_{\mu\nu}
      \ ,
\end{equation}
where we have imposed the boundary condition  
$g^{(1)}_{\mu\nu} (y=0, x^\mu ) =0 $. 

\subsection{Second order}

If we ignore the $\chi$ field then we get the following results:
\begin{equation}
  K^{(2)}  = {l^3 \over 2(d-2)^2 a^4} 
        \left( R^\alpha_{\ \beta} R^\beta_{\ \alpha} 
            - {3d-4\over 4(d-1)^2}  R^2 \right) -{l^3 \over 2(d-1)(d-2) a^2} 
            \left( R^\alpha_{\ \beta} R^\beta_{\ \alpha} 
            - {1\over 2(d-1)}  R^2 \right)   \ .
\end{equation}
\begin{eqnarray}
  \Sigma^{(2)\mu}_{\ \ \  \nu} &=& 
        {l^3\over (d-2)(d-4) a^4}
                \left[ R^\mu_{\ \alpha} R^\alpha_{\ \nu} 
                - {1\over d-1} R R^\mu_{\ \nu}
                -{1\over 2} ( R^{\alpha\mu}_{\ \ |\nu\alpha} + 
                        R^{\alpha \ |\mu}_{\nu \ | \alpha} ) 
                        +{1\over 2} \Box R^\mu_{\ \nu}
                        \right. 
                        \nonumber \\
                & & \quad 
                \left. 
                +{d\over 4(d-1)} R^{|\mu}_{\ |\nu}
                -{1\over 4(d-1)}\delta^\mu_\nu \Box R 
                -{1\over d} \delta^\mu_\nu R^\alpha_{\ \beta} 
                        R^\beta_{\ \alpha}
                +{1\over d(d-1)} \delta^\mu_\nu R^2 
                \right] + {l^3 \over a^d}t^\mu_\nu \nonumber \\
                & & \quad
                -{l^3\over (d-2)^2 a^2}
                \left[ R^\mu_{\ \alpha} R^\alpha_{\ \nu} 
                        - {1\over 2(d-1)} R R^\mu_{\ \nu}
                -{1\over 2} ( R^{\alpha\mu}_{\ \ |\nu\alpha} 
                + R^{\alpha \ |\mu}_{\nu \ | \alpha} ) 
                +{1\over 2} \Box R^\mu_{\ \nu}
                \right. \nonumber \\
                & & \quad 
                \left.
                +{d\over 4(d-1)}R^{|\mu}_{\ |\nu} 
                -{1\over 4(d-1)}\delta^\mu_\nu \Box R 
                        -{1\over d} \delta^\mu_\nu R^\alpha_{\ \beta} 
                        R^\beta_{\ \alpha} 
                +{1\over 2d(d-1)} \delta^\mu_\nu R^2 
                \right] \ ,
\end{eqnarray}
where the homogeneous solution $t^\mu_{\ \nu} $ satisfies the transverse
and traceless conditions 
\begin{equation}
  t^\mu_{\ \nu | \mu} =0  \ ,  \quad   t^\mu_{\ \mu} = 0   \ .
\end{equation}
This is the point where the dependence on the dimensions appears.
 We do not have a trace anomaly, in contrast to the case of $d=4$. 

\subsection{Effective Equations and Effective Action}

The consequences of the junction condition (10), order by order, are the 
following. 
 At   zeroth order, we have
\begin{equation}
   \left[ K^{(0)\mu}_\nu - \delta^\mu_\nu K^{(0)} \right] |_{y=0}
    = -{d-1 \over l} 
    = - {\kappa^2 \over 2} \sigma \delta^\mu_\nu   \ .
\end{equation}
 Thus, we get the relation, $\kappa^2 \sigma = 2(d-1)/l$ and assume that this 
relation holds exactly.

 At   first order, we obtain
\begin{equation}
  \left[ K^{(1)\mu}_{\quad\  \nu} - \delta^\mu_\nu K^{(1)} \right] |_{y=0}
  = {l\over d-2} \left( R^\mu_\nu -{1\over 2} \delta^\mu_\nu R \right) 
  + \chi^\mu_\nu
  = {\kappa^2 \over 2} T^\mu_\nu  \ .
\end{equation}
 The homogeneous solution $\chi_{\mu\nu}$ is the generalized dark radiation.
 Supposing the relation $8\pi G_N = (d-2)\kappa^2 / 2l $ holds, then
 the conventional Einstein equation can be recovered on the brane at this order
 when we put $\chi_{\mu\nu} =0 $. This can be performed without losing 
 consistency.
   
 Finally, up to  second order, the junction condition gives
\begin{eqnarray}
 {l\over d-2} ( R^\mu_{\ \nu} &-& {1\over 2} \delta^\mu_\nu R ) 
        + {l^3 \over (d-2)^2(d-4)} 
        \left[ 2R^\mu_{\ \alpha} R^\alpha_{\ \nu}
        -{d\over 2(d-1)}R R^\mu_{\ \nu}
        -( R^{\alpha\mu}_{\ \ |\nu\alpha} + R^{\alpha \ |\mu}_{\nu \ | \alpha})
        +\Box R^\mu_{\ \nu} \right. \nonumber \\
        &+&  
        \left. {d\over 2(d-1)} R^{|\mu}_{\ |\nu}
        -{1\over 2(d-1)}\delta^\mu_\nu \Box R  
        -{1\over 2} \delta^\mu_\nu R^\alpha_{\ \beta} R^\beta_{\ \alpha}
        +{d\over 8(d-1)} \delta^\mu_\nu R^2 
         \right] 
        = {\kappa^2 \over 2} T^\mu_{\ \nu}    \ .
\end{eqnarray}
We calculate the projective Weyl tensor to find the geometrical meaning of
the above equation as
\begin{eqnarray}
  E^{(2)\mu}_{\quad\ \nu} 
        &=& {dl^2\over (d-2)^2(d-4)}  
                \left[
                R^\mu_{\ \alpha} R^\alpha_{\ \nu}
                -{1\over d-1} RR^\mu_{\ \nu} 
                -{d-2\over d}( R^{\alpha\mu}_{\ \ |\nu\alpha} + 
                R^{\alpha \ |\mu}_{\nu \ | \alpha} )
                +{d-2\over d} \Box R^\mu_{\ \nu} \right. \nonumber \\
                &+& 
                \left. {d-2\over 2(d-1)} R^{|\mu}_{\ |\nu}
                -{d-2\over 2d(d-1)} \delta^\mu_\nu \Box R 
                -{1\over d} 
                        \delta^\mu_\nu R^\alpha_{\ \beta} R^\beta_{\ \alpha} 
                +{1\over d(d-1)}\delta^\mu_\nu R^2 
                \right]   \ .
\end{eqnarray}
Substituting this expression into Eq.(A15) yields our main result
\begin{equation}
  G^{(4)}_{\mu\nu} = {(d-2)\kappa^2 \over 2l} T_{\mu\nu} 
  +l^2 P_{\mu\nu}   -E^{(2)}_{\mu\nu}  \ ,
\end{equation}
where
\begin{equation}
     P^\mu_{\ \nu } =  -{1\over (d-2)^2}
                \left[ 
                R^\mu_{\ \lambda} R^\lambda_{\ \nu}  
                -{d\over 2(d-1)}R R^\mu_{\ \nu} 
                -{1\over 2}\delta^\mu_\nu R^\alpha_{\ \beta} R^\beta_{\ \alpha}
                +{d+2\over 8(d-1)} \delta^\mu_{\ \nu} R^2
                \right]  \ .
\end{equation}
Within the accuracy we are considering, using the lowest order equation 
$R^\mu_{\ \nu} 
= {\kappa^2 /l} [(d-2)/2T^\mu_{\ \nu} - 1/2 \delta^\mu_\nu T ] $, 
we get formally the same result as that of Shiromizu et al.: 
\begin{equation}
  G^{(4)}_{\mu\nu} = 8\pi G_N T_{\mu\nu} + \kappa^4 \pi_{\mu\nu} 
       - E^{(2)}_{\mu\nu}  \ ,
\end{equation}
with
\begin{equation}
  \pi_{\mu\nu} = -{1\over 4}T_\mu^\lambda T_{\lambda\nu} 
  +{1\over 4(d-1)}TT_{\mu\nu} + {1\over 8} g_{\mu\nu} \left( 
   T^{\alpha\beta} T_{\alpha\beta} -{1\over d-1} T^2 \right)   \ .
\end{equation}
Thus, we have established the correspondence between the geometrical and 
AdS/CFT approaches in any dimensions.


\end{document}